\documentclass[%
reprint,
showpacs,preprintnumbers,
 amsmath,amssymb,
aps,
prd,
twocolumn
]{revtex4}

\usepackage{dcolumn}
\usepackage{bm}
\usepackage{longtable}

\newcommand{\yhe}{\ensuremath{Y_\mathrm{p}}}
\newcommand{\obh}{\ensuremath{\Omega_bh^2}}
\newcommand{\neff}{\ensuremath{N_\mathrm{eff}}}
\newcommand{\nnu}{\ensuremath{N_\nu}}
\newcommand{\apjs}{Astrophys. J. Suppl.}
\newcommand{\apjl}{Astrophys. J. Lett.}
\newcommand{\mnras}{Mon. Not. Roy. Astron. Soc.}
\newcommand{\jcap}{J. Cosmo. Astropart. Phys.}
\newcommand{\physrep}{Phys. Rep.}
\newcommand{\aap}{Astron. Astrophys.}
\newcommand{\ssr}{Space Sci. Rev.}
\newcommand{\araa}{Ann. Rev. Astron. Astrophys.}

\usepackage{graphicx}

\begin{document}


\title{An analysis of constraints on relativistic species from primordial nucleosynthesis and
the cosmic microwave background}

\author{Kenneth M. Nollett}
 \email{nollett@anl.gov}
\affiliation{%
Physics Division, Argonne National Laboratory, Argonne, IL~~60439, USA
}%
\author{Gilbert P. Holder}
\affiliation{Department of Physics, McGill University, Montreal, QC H3A 2T8,
Canada}

\date{\today}

\begin{abstract}

We present constraints on the number of relativistic species from a
joint analysis of cosmic microwave background (CMB) fluctuations 
and light element abundances (helium and deuterium) compared to
big bang nucleosynthesis (BBN) predictions.  Our BBN
calculations include updates of nuclear rates in light of recent
experimental and theoretical information, with the most significant
change occuring for the $d(p,\gamma)^3\mathrm{He}$ cross section.  We
calculate a likelihood function for BBN theory and observations that
accounts for both observational errors and nuclear rate uncertainties
and can be easily embedded in cosmological parameter fitting. We then
demonstrate that CMB and BBN are in good agreement, suggesting that
the number of relativistic species did not change between the time of
BBN and the time of recombination.  The level of agreement between BBN
and CMB, as well as the agreement with the standard model of particle
physics, depends somewhat on systematic differences among
determinations of the primordial helium abundance. We demonstrate that
interesting constraints can be derived combining only CMB and D/H
observations with BBN theory, suggesting that an improved D/H
constraint would be an extremely valuable probe of cosmology.

\end{abstract}


\maketitle

\section{Introduction}

The sensitivity of early-universe cosmology to the number and
properties of light particle species was recognized early in the
development of the hot big bang model
\cite{hayashi50,alpher53,peebles66,shvartsman69}.  The expansion rate
of the universe at early times increases with number of relativistic
particle species in thermal equilibrium, and this in turn sets
timescales for big bang nucleosynthesis (BBN).  One can then use the
BBN yields of light nuclei to constrain the number of light species
quantitatively \cite{steigman77}.  Constraints on the expansion rate
are therefore often expressed in units of an effective number of light
fermion species \neff, so that for the standard model of particle
physics $\neff \sim 3$, while $\neff\neq 3$ corresponds to some
altered expansion rate that might be caused by additional neutrino
species or by some other effect (like a nonzero neutrino chemical
potential or violations of general relativity) \footnote{Even within
  the standard model, $\neff\neq 3$ after the temperature drops below
  the electron rest mass.  At that time, all electron-positron pairs
  annihilate, transferring their energy mainly into the photon bath.
  A small fraction of annihilations produce neutrino pairs that fail
  to thermalize, contributing energy equivalent to 0.046 of a
  relativistic species to the subsequent evolution \cite{mangano05}.
  In the CMB literature, \neff\ within the standard model is generally
  taken to be 3.046, but since pair annihilation occurs during BBN,
  most BBN work refers to the standard-model case as $\neff=3$.  We
  report results in terms of \neff\ at late times, so that the
  standard model has $\neff=3.046$.}.  Recent BBN-derived constraints
may be found in Refs. \cite{steigman10,hamann10,mangano11}.  The
cosmic microwave background (CMB) is sensitive to the expansion rate
much later: at the times of equal matter and radiation densities and
of hydrogen recombination.  Recent measurements of the CMB power
spectrum now allow interesting constraints on
\neff\ at those times \cite{wmap7,hamann11-1,keisler11,hou11}.

Constraints on light species from BBN and the CMB have been examined
many times as both observational data and theoretical calculations
have improved.  The consistency or not of these two types of
constraints is an important question, particularly since they probe
expansion rates at widely separated times in the evolution of the
universe.  In the present work, we check for consistency between the
two data sets, examine the sensitivity of that consistency to model
assumptions, and then attempt to place the strongest constraint
possible on \neff\ by combining the two types of data.

The BBN yields -- especially of deuterium -- depend not just on
\neff\ but also on the cosmological baryon density, which in units of
the critical density of the universe is $\Omega_b\equiv 8 \pi G
\rho_B/3h^2$.  ($h$ is the Hubble constant in units of 100
$\mathrm{km\,s^{-1}\,Mpc^{-1}}$, $G$ is Newton's constant, and
$\rho_B$ is the baryon density expressed as mass per unit volume.)
Prior to the precise inference of $\obh$ from CMB measurements, the
strongest constraint on \neff\ came from using the
deuterium-to-hydrogen number ratio D/H to restrict $\obh$ and then
exploiting the \neff\ dependence of the primordial helium mass
fraction \yhe.  However, D/H has its own dependence on
\neff\ \cite{cardall96,steigman10}; a strong external constraint on
$\obh$ allows BBN constraints on \neff\ that are independent of \yhe.
We show below that, particularly because precise measurements of
\yhe\ are difficult, the constraint on \neff\ from D/H is competitive
with that from \yhe.

In this paper, we focus on three issues: a full update of the BBN
reaction rates including a revision to the $d(p,\gamma)^3\mathrm{He}$
rate based on recent theoretical calculations; an implementation of
BBN constraints that includes both observational and theoretical
uncertainties within CosmoMC \cite{lewis02} (a popular Markov chain
Monte Carlo fitting program used for cosmological data); and the
utility of D/H as a probe of new physics that is independent of
uncertainties in the primordial helium abundance.

\section{Big-Bang Nucleosynthesis and \neff}

BBN constraints on physics and cosmology require accuracy in both
computed BBN yields and observed abundances.  With only standard-model
physics (\nnu=3, no late-decaying particles, negligible neutrino
chemical potentials, etc.), BBN is a one-parameter model, depending on
the baryon density.  Since $\obh=0.02249(55)$ \cite{wmap7} is
now known to $\sim 3\%$ from the CMB alone, this ``standard'' BBN is a
high-precision theory with no free parameters. The predicted \yhe\ has
a precision of $\sim 0.2\%$, while yields of D, $^3$He, and $^7$Li are
predicted with respective precisions of roughly $5\%$, $4\%$, and
$8\%$ (computed below and e.g. in
Refs.\cite{cyburt04,serpico04,nollett00,coc04,cyburt08}).  To
constrain deviations from this standard model, we need both accurate
BBN calculations based on up-to-date nuclear inputs and
observationally-inferred primordial abundances.

\subsection{The BBN calculation}

We compute BBN light-nuclide yields using a modified version of the
Wagoner/Kawano code described in Ref. \cite{kawano92}.  We adjusted
the timestep controls 
to compute yields more precisely than 0.1\% for given nuclear rates,
and we used the Monte Carlo techniques described in
Ref. \cite{nollett00} to propagate nuclear cross section data into
yields and yield uncertainties.  For the centroid of our
\yhe\ predictions we used the independent BBN code of Lopez and Turner
\cite{lopez99}.  It accounts for several small effects in
weak-interaction physics and the equation of state and achieves a
theoretical precision (for a given neutron mean lifetime $\tau_n$) of
$\Delta \yhe < 0.0002$ or $< 0.1\%$.  The \yhe\ from this code has
been shown to agree within this margin with somewhat independent
calculations \cite{olive00}.  We note that its results are $\Delta
\yhe \sim 0.0004$ higher than those reported in Ref. \cite{serpico04}
for reasons that are unclear -- each calculation should have a higher
precision than this.  For comparison, the uncertainty arising from the
currently recommended $\tau_n$ is $\sim 0.0003$ (but see below for its
likely underestimation).

BBN codes integrate a set of coupled rate equations to evolve
light-nuclide abundances in the expanding and cooling universe.  The
rate coefficients in those equations are in turn based on nuclear
cross sections, mainly inferred from experiment.  Several improvements
in the nuclear inputs have occurred in the decade since the study of
Ref. \cite{nollett00}, so we have updated the input database used
there accordingly.  We now summarize those changes with an emphasis on
the D and $^4$He yields that are useful for constraining \neff.  In
addition to the reactions affecting \yhe\ and D/H, the input database
for the reaction $^3\mathrm{He}(\alpha,\gamma)^7\mathrm{Be}$ has also
been brought up to date by including recent data examined in
Refs. \cite{cyburt08,adelberger11}; this affects only the curves shown
in Fig. \ref{fig:schramm}
and almost none of the discussion.

\subsubsection{The reaction $p(n,\gamma)d$}

There are very few cross section data for neutron capture on protons
at energies relevant for BBN, though the situation has improved in
recent years \cite{hara03}, and their precision is generally rather
low.  For this reason, BBN evaluations in the 1990s
\cite{smith93,nollett00} used theoretical cross sections from the
ENDF/B-VI and earlier databases \cite{hale97} rather than fitting the
data.  In general, theoretical calculations more precise than the data
at BBN energies \cite{hale97,carlson98,marcucci06} were possible, but
some calculations had only been carried out for thermal neutrons, and
uncertainties were poorly quantified.

The BBN calculations of Refs. \cite{smith93,nollett00} applied to the
ENDF cross section a conservative 5\% error in overall normalization
that had a somewhat murky origin in the literature.  It has
subsequently been shown using pionless effective field theory (EFT)
that the low-energy cross section can be computed from a small number
of input quantities (most at least implicitly used in the ENDF cross
section) and shown to have an error of $< 1\%$ at BBN energies
\cite{rupak00,ando06}.  We have accordingly used the EFT results
\cite{rupak00} to compute the rate of $p(n,\gamma)d$, assigned it an
overall normalization error of $1\%$, and incorporated the results
into the BBN code in place of the ENDF/B-VI rate.  Since the EFT rate
agrees with the old ENDF rate, this results in a much smaller error on
D/H but does not shift the most-likely yield.

\subsubsection{The reactions $d(d,n)^3\mathrm{He}$ and $d(d,p)^3\mathrm{H}$} 

The investigations in Ref. \cite{nollett00} resulted in a call for
improved data on deuteron-deuteron reactions at 100 to a few hundred
keV (center-of-mass), where there were no modern data with
well-quantified errors.  The response was the work of
Ref. \cite{leonard06}, which produced data with a precision of $\sim
2\%$.  We added these results to our rate database along with
lower-energy data \cite{greife95} that were missed in the literature
search of Ref. \cite{nollett00}.  The addition of these data reduces
the error on D/H, particularly at low values of $\obh$ where
$d+d$ reactions formerly dominated the error budget \cite{nollett00}.
The new data shift D/H by less than 1\% relative to previous results
\cite{nollett00,burles01}.

\subsubsection{The reaction $d(p,\gamma)^3\mathrm{He}$}

It was found in Ref. \cite{nollett00} and by other authors
\cite{fiorentini98,cyburt04,coc10} that the only reactions important
for the uncertainty on the D/H yield were the three reactions
discussed above and proton capture on deuterium.  Even before the
above updates to $p(n,\gamma)d$ and $d+d$ rates, the
$d(p,\gamma)^3\mathrm{He}$ rate dominated the error on D/H at the
CMB-inferred $\obh$.

\begin{figure}
\includegraphics[width=3.3in]{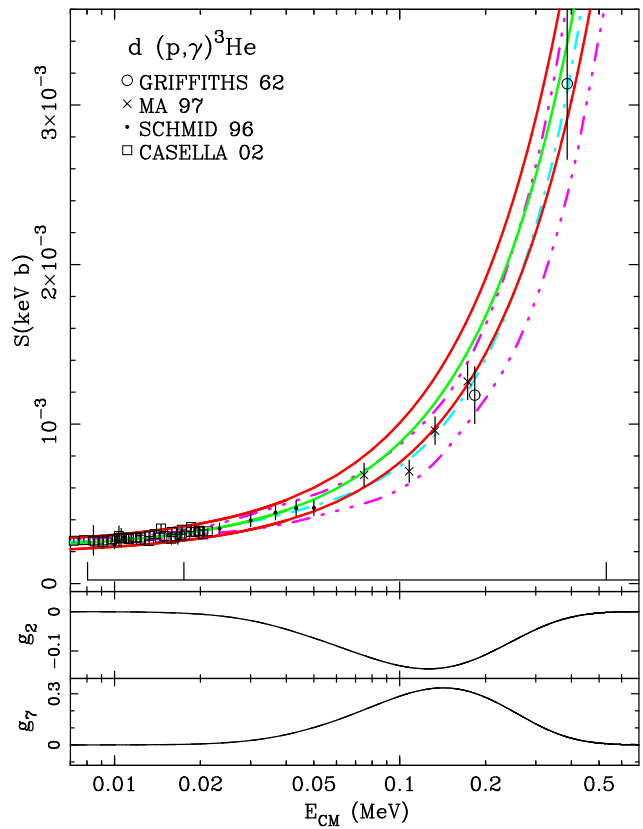}  
  \caption{(Color online) Upper Panel: $S$-factor data for the reaction
    $d(p,\gamma)^3\mathrm{He}$ are shown, along with the best-fit
    curve and $2\sigma$ intervals based purely on those data
    \cite{nollett00} (dash-dot and dash-dot-dot-dot curves) and on the
    \textit{ab initio} calculation \cite{viviani00,marcucci05}
    (solid).  All errors are shown as $2\sigma$ intervals for
    comparability with Fig. 3 of Ref. \cite{nollett00}.  Lower Panels:
    Sensitivity functions \cite{nollett00} showing the relative
    importance of this reaction at varying energy for D/H ($g_2$) and
    Li/H ($g_7$) yields, evaluated at $\obh=0.019$.  Convolving
    the $S$-factor error band with the sensitivity function gives the
    total uncertainty arising from this reaction.  The strongest
    sensitivity to the $d(p,\gamma)^3\mathrm{He}$ rate occurs at the
    largest $|g_i|$, in the 50--500 keV range.}
  \label{fig:dpgamma}
\end{figure}

The effect of using measured $d(p,\gamma)^3\mathrm{He}$ cross sections
in BBN calculations has not changed since our previous study
Ref.~\cite{nollett00}.  There has been one published low-energy cross
section measurement since then \cite{luna02}, and we have updated our
database to include it.  As the new data are well below most of the
BBN energy window shown in Fig. \ref{fig:dpgamma}, and they agree with
previous data there, they do not affect our results discernably.  This
calculation will be referred to as ``Empirical dp$\gamma$'' below.

Fig. \ref{fig:dpgamma} shows the scarcity of laboratory data for this
reaction at BBN energies that was cited in Ref. \cite{nollett00} as a
cause for concern in both rate determination and error estimation.
Cross section fitting for charged-particle reactions is generally done
with $s$-wave Coulomb barrier penetration divided out, in terms of the
$S$-factor
\[
S(E) = Ee^{2\pi Z_1Z_2\alpha c/v}\sigma(E),
\]
where $\sigma$ is the cross section, $E$ the energy, $Z_i$ the nuclear
charges, $\alpha$ the fine-structure constant, $c$ the speed of light,
and $v$ the relative velocity.  Except for one rate evaluation using
the $R$-matrix formalism \cite{descouvemont04}, all BBN work to date
has fitted $S(E)$ either as a polynomial with poor physical motivation
\cite{smith93,cyburt04,serpico04} or as an expansion in
linearly-independent, locally-supported $B$-splines \cite{nollett00}
in which $S(E)$ at a given energy is determined by nearby data.  There
is only one modern data set for $d(p,\gamma)^3\mathrm{He}$ at the most
important energy range of 50 -- 500 keV \cite{ma97}, so only the
$R$-matrix approach allows independent measurements to average against
it in a physically motivated way.

Because \textit{ab initio} nuclear theory is now quite accurate for
many bound-state and continuum properties of three-body systems
\cite{kievsky08}, a purely empirical approach to
$d(p,\gamma)^3\mathrm{He}$ is no longer required; the situation with
$d(p,\gamma)^3\mathrm{He}$ is now similar to that with $p(n,\gamma)d$.
In particular, the Pisa group has computed the cross section of
$d(p,\gamma)^3\mathrm{He}$ using accurate two- and three-nucleon
potentials, corresponding two- and three-body corrections to the
electromagnetic currents, and accurate methods of solution
\cite{viviani00,marcucci05}.  The results are in very good agreement
with data above and below the BBN energy range (particularly the new
low-energy data of Ref. \cite{luna02}), not only in total cross
section but also in angular-distribution and polarization observables.
However, the total cross section is well above most of the sparse data
at BBN energies.

The energy dependence of the cross section is determined by
competition between $s$- and $p$-wave capture, which is probed by the
polarization observables.  Since both they and the low-energy total
cross sections support the model, we suspect that the data of
Ref. \cite{ma97} may have a systematically low normalization by about
20\% (compared with their stated systematic error of 9\%) and that the
theoretical curve provides a better guide to the actual cross section.

For these reasons, we adopt the \textit{ab initio} cross section
\cite{viviani00,marcucci05} for our BBN calculations.  The precision
of the calculation is likely a few percent at worst, but it is
difficult to determine.  Errors could in principle be propagated
through the constraints on ``model-dependent'' two-body currents, but
these are themselves small corrections to the main two-body
corrections that make up $\sim 15\%$ of the cross section.  The
remaining errors arise from the numerical method of computation
(probably negligible) and the nuclear interaction model.  The methods
that have been used to construct the modern nucleon-nucleon potentials
($\sim 40$ parameters fitted to thousands of data points from many
experiments) have not been adapted to facilitate error propagation.  
The modern potentials do reproduce many properties of
two- and three-nucleon systems \cite{wiringa95,marcucci06,kievsky08}
at the percent level, and they have also been applied very
successfully to larger systems (e.g., \cite{pieper08}).

Because assigning a quantitative error to the computed cross section
is difficult, we would like to be conservative.  The case of
$p(n,\gamma)d$ is again analogous: through the 1990s, errors of 5\%
were assigned to an evaluation that later proved accurate to something
more like 1\% \cite{smith93,nollett00}.  The review of
Ref. \cite{adelberger11} presents a polynomial fit to the
$d(p,\gamma)^3\mathrm{He}$ $S$-factor and a recommended cross section
primarily for use well below the BBN range.  This fit's normalization
is dominated by the precise and copious low-energy data, which cause
the recommended $S(0)$ to have an error of 7\%.  The \textit{ab
  initio} cross section is in essentially perfect agreement with the
low-energy data, so we adopt 7\% as the precision to which it has been
tested and apply this as an overall error on its normalization.  This
error is handled in our BBN Monte Carlo procedure just like that for
the $p(n,\gamma)d$ cross section.  Ref. \cite{nollett00} in effect
propagated a 9\% error bar from the nuclear data, so adopting 7\% does
not drastically change the (dominant) contribution of
$d(p,\gamma)^3\mathrm{He}$ to the yield errors.  We use these results
rather than the ``Empirical dp$\gamma$'' calculation described above
in our likelihood analysis.

The larger \textit{ab initio} cross section for
$d(p,\gamma)^3\mathrm{He}$ at BBN energies results in more destruction
of deuterium at late times than was found previously
\cite{esmailzadeh91,nollett00}.  At the WMAP baryon density \citep{wmap7}, this
amounts to about a 5.6\% reduction of D/H; the D/H-inferred
$\obh$ is shifted downward by roughly half a standard
deviation.  In light of the severity of this shift, independent cross
section measurements at 50--500 keV would be quite useful.  The shift
in the $d(p,\gamma)^3\mathrm{He}$ rate also increases the $^7$Li yield
by about 10\% through its effect on the availability of $^3$He at late
times, worsening slightly the primordial $^7$Li problem reviewed in
Ref. \cite{fields11}.

\subsubsection{The neutron lifetime $\tau_n$}

The mean lifetime $\tau_n$ of the neutron is used in the BBN
calculation to normalize all of the weak-interaction rates that
interconvert neutrons and protons at early times.  It may be
conveniently thought of as determining the weak coupling constant
$G_F$, and its uncertainties propagate through the weak rates into the
abundance yields.

The work of Refs. \cite{nollett00,lopez99} used a neutron mean
lifetime of $\tau_n=885.4\pm 2.0$ s, which was the most recent
experimental result at the time.  This result was subsequently
included in the world average of $885.7\pm 0.8$ s recommended for
several years by the Particle Data Group \cite{pdg02}.  More recently,
conflicting lifetimes of $878.5 \pm 0.7 \pm 0.3$ s and $880.7\pm
1.3\pm 1.2$ s have been reported respectively by Serebrov et
al.~\cite{serebrov05,serebrov08} and Pichlmaier et
al.~\cite{pichlmaier10}.  The Particle Data Group now recommends a
world average that includes conflicting values, $881.5 \pm 1.5$ s
\cite{pdg10}, with errors that have been inflated to reflect the
discrepancy.

We performed our calculations using the currently recommended $\tau_n$
with its recommended error, and these form the basis of our reported
results.  We also performed BBN calculations with the formerly
recommended $\tau_n = 885.7\pm 0.8$ s and with $\tau_n = 878.5\pm 1$ s
from Serebrov.  The difference between the two amounts to $\Delta \yhe
\simeq 0.0015$, equivalent to $\Delta\neff \simeq 0.12$.  The effect on
D/H is a shift of $0.4\%$ at $\obh=0.0225$.  For both yields,
the difference is much smaller than the other uncertainties in our
analysis.

\subsection{The \neff\ dependences of $\mathrm{D/H}$ and \yhe}

\begin{figure*}

\includegraphics[width=3.3in]{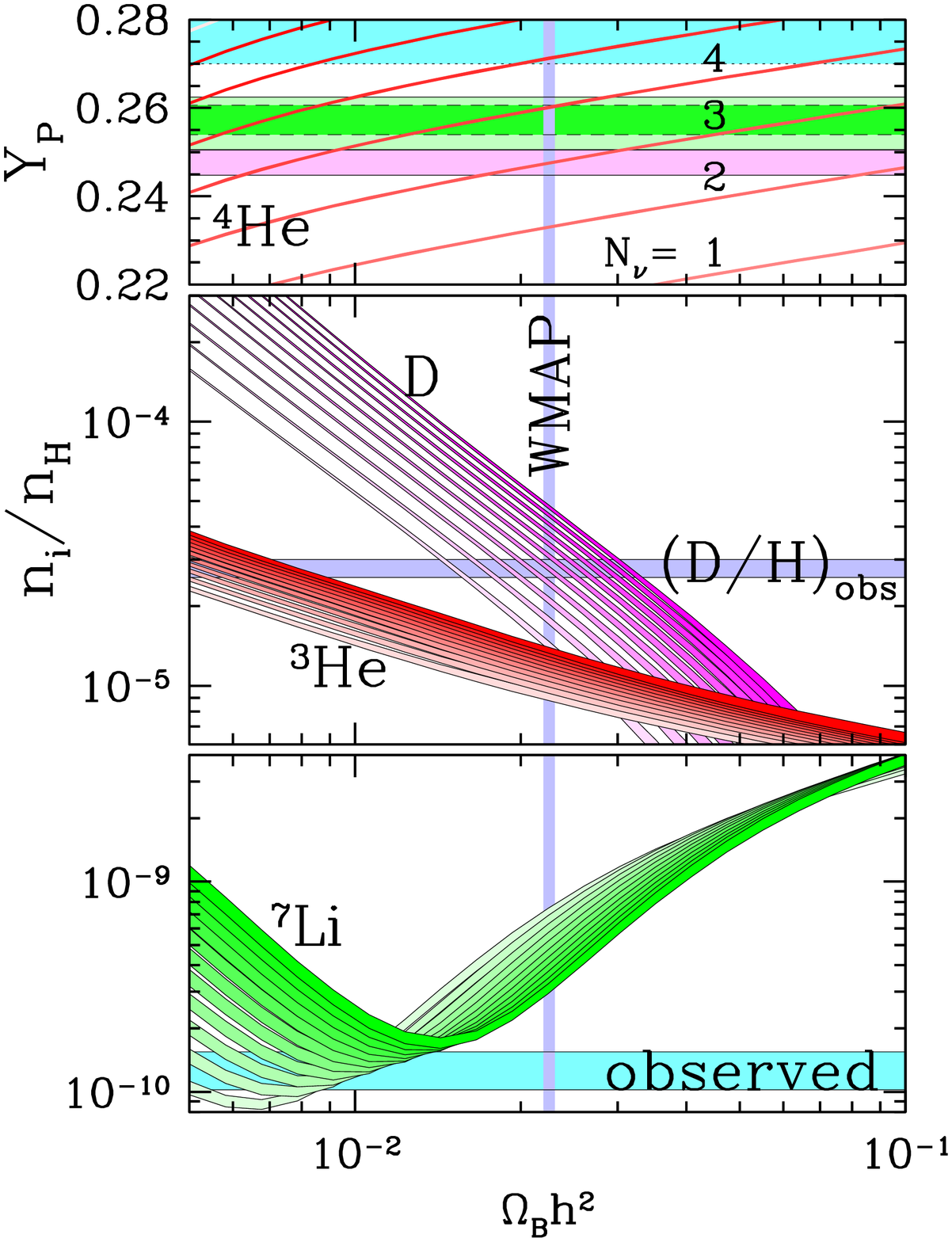}  
\includegraphics[width=3.1in]{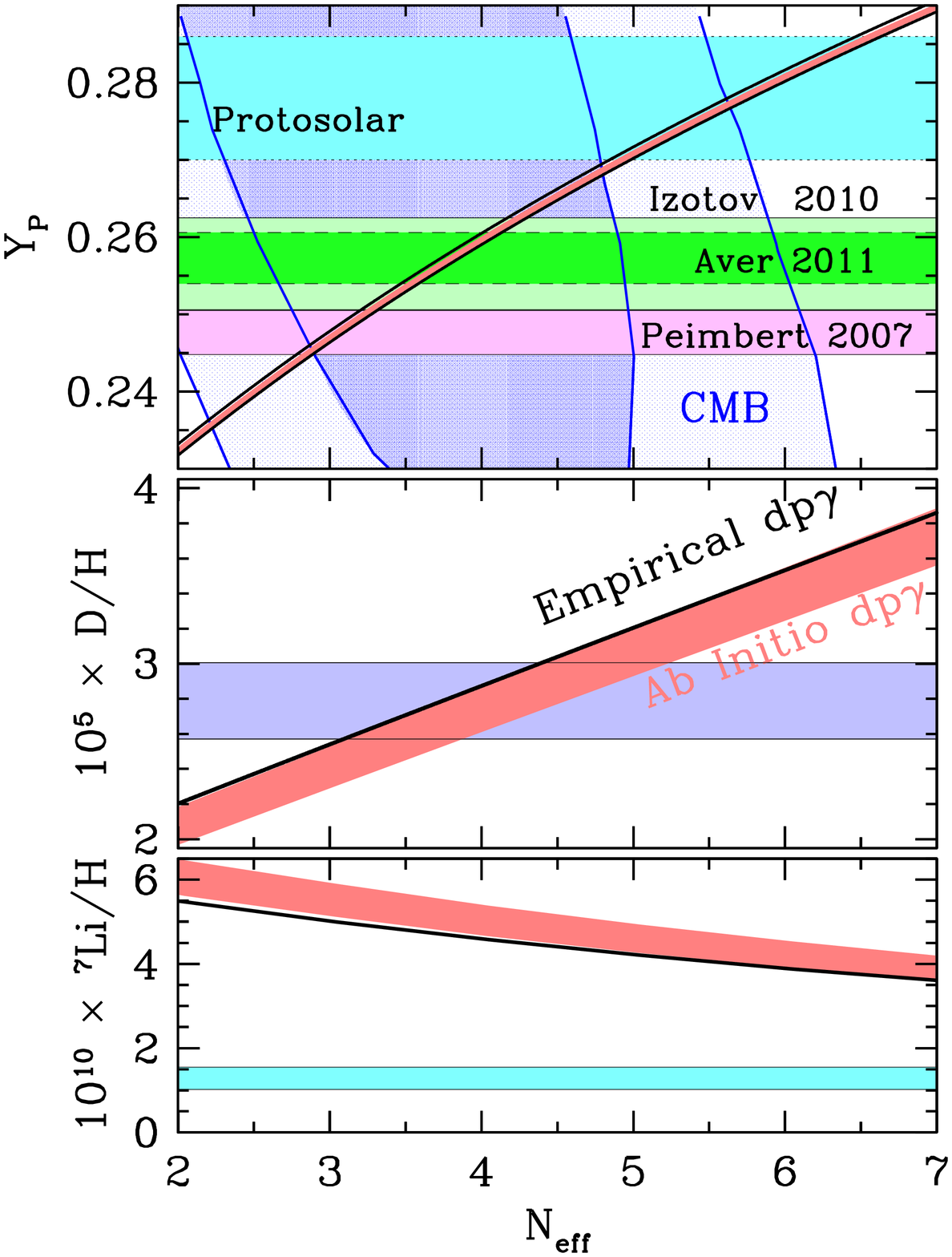}  
  \caption{(Color online) Left panels: Yields of light nuclides as
    functions of the baryon density, $\obh$.  Each band
    indicates yields for a single value (integer plus 0.046) of
    \neff\ and its thickness the $1\sigma$ nuclear uncertainty on
    those yields.  Bands are shaded lighter for smaller values of
    \neff\ and darker for larger \neff.  Also shown are horizontal
    bands indicating observational constraints on abundances
    \cite{peimbert07,bonifacio07,pettini08,izotov10,aver11} and a
    vertical band indicating the WMAP7 $1\sigma$ interval for
    $\obh$.  Right panels: The same yields shown as functions
    of \neff\ at the CMB-inferred value of \obh, essentially slices
    along the band labeled ``WMAP'' in the left panels.
    Salmon-colored bands indicate yields with our adopted rates, and
    their widths indicate a quadrature sum of nuclear errors and the
    error on \obh.  In the \yhe\  graph, blue $1\sigma$ and $2\sigma$
    contours indicate constraints from WMAP7+SPT alone, while black
    curves paralleling the adopted yields indicate the effects of
    using the old Particle Data Group \cite{pdg02} (upper) and Serebrov
    \cite{serebrov08} (lower) values of $\tau_n$ (no errors shown).
    In the D/H and $^7$Li/H graphs, the thin solid curves indicate the
    result of using the empirical rate for $d(p,\gamma)^3\mathrm{He}$,
    again without showing the errors.  Horizontal bands indicate
    observational constraints as discussed in the text.}
  \label{fig:schramm}
  \label{fig:bbn-variants}
\end{figure*}

We computed BBN yields, including their errors and correlation
coefficients, on a grid covering $\obh$ from 0.005 to 0.1 and
\neff\ from 0.046 to 10.046.  This includes the entire range sampled
in our Markov chain Monte Carlo (MCMC) analysis below, allowing
considerable margins at the edge of the grid.  We show all BBN yields
as functions of $\obh$ over this grid in the left panels of
Fig.~\ref{fig:schramm}.  Separate curves are shown, with widths
indicating $1\sigma$ errors, at integer intervals in \neff.  This
indicates the nature and strength of the sensitivity to \neff.  The
dependence of \yhe\ and D/H on \neff\ is apparent; that of \yhe\ is
much larger, relative to its yield uncertainties, than that of D/H.
At the band indicating the current WMAP value of $\obh$, D/H changes
by about 10\% per unit \neff.

The right side of Fig. \ref{fig:bbn-variants} shows the dependences of
BBN yields on \neff\ and on the assumed rates of
$d(p,\gamma)^3\mathrm{He}$ and of the weak processes scaled by
$\tau_n$.  Each yield was computed at $\obh=0.02260(54)$, inferred
below from a cosmological analysis of CMB fluctuations marginalized
over \yhe\ and \neff, and the error on $\obh$ was added in quadrature
to the nuclear uncertainties.  The diagonal curves are thus
essentially slices along the ``WMAP'' band in the left panels of
Fig. \ref{fig:schramm}.  The effects on D/H and $^7$Li/H of switching
$d(p,\gamma)^3\mathrm{He}$ rates are shown as black curves with no
error band in the right side of Fig. \ref{fig:bbn-variants}; that on
D/H corresponds roughly to $\Delta\neff=0.42$.  Similarly, the effects
of assuming either the pre-2011 or the Serebrov values of $\tau_n$ are
shown as black curves without error bands next to the \yhe\ yield
curve.  To a very good approximation, $d(p,\gamma)^3\mathrm{He}$ does
not affect \yhe\ and $\tau_n$ does not affect D/H.

The yields \yhe\ and D/H measure slightly different things
\cite{steigman77}.  At the start of standard BBN the population of
relativistic particles consists of photons, electrons, positrons, and
three species of neutrinos, all at the same temperature.  At a given
temperature, the resulting expansion rate of the universe is $2.3$
times that from photons alone.  Weak freezeout begins at this time,
setting the neutron/proton ratio that will eventually determine \yhe.
An additional neutrino species speeds the expansion rate by $\sim
40.3\%$, so that weak freezeout occurs at higher temperature.  Higher
temperature at equilibrium implies more neutrons, so faster expansion
yields more $^4$He.

In contrast, the abundance of deuterium is determined by $d+d$ and
$d+p$ reactions toward the end of BBN, when the photon temperature is
well below the electron rest mass.  At these times, there are
essentially no electrons or positrons, and the energy they previously
carried has been transferred to the photons, which are consequently at
higher temperature than the neutrinos.  The expansion rate at this
time in standard BBN is $\sim 1.7$ times that of photons alone, so
that adding a neutrino species at the same temperature as the others
causes a $\sim 36.7\%$ speed-up.  Faster expansion implies less time
available for deuterium burning and thus higher D/H.  In the standard
model, deuterium burning and the recombination era share the same
populations of relativistic species: photons plus three neutrino
species at $(4/11)^{1/3}$ of the photon temperature.  


\subsection{Observed primordial abundances}

Constraints based on BBN require observed abundances.  The inference
of each primordial abundance is a highly technical subject unto itself
with its own extensive literature.  Here we provide brief descriptions
of the experimental constraints and explain the specific values that
we choose.  The interested reader is referred to the cited literature
for more information, and to the reviews in
Refs. \cite{steigman07,simha08,pdg10,iocco09,steigman10,fields11}.

\subsubsection{$^4$He}

The primordial $^4$He mass fraction \yhe\ is inferred from
emission-line spectra of H{\sc ii} regions in metal-poor blue compact
galaxies.  Lines of H and He are observed and a He/H ratio inferred
using a radiative transfer model; the result is customarily expressed
in terms of \yhe.  Because of the weak dependence of \yhe\ on
$\obh$ evident in Fig. \ref{fig:schramm},
the desired precision of measurement is in the percent range.  Large
data sets have been assembled, and the errors on \yhe\ are dominated
by systematic effects in the analysis, not statistics.

A recent revision of the atomic data used to infer \yhe\
\cite{porter05} resulted in a systematic upward shift of all data sets
(but not all individual data).  Three groups have published \yhe\
values based on the new emissivities, and we mention only the most
recent publication from each.  Peimbert et al. \cite{peimbert07} find
$0.2477\pm 0.0029$ from 5 H{\sc ii} regions using radiative-transfer
models tailored to each object and taking into account temperature
fluctuations within each H{\sc ii} region.  Applying approaches in
which physical parameters for each H{\sc ii} region are determined in
a simultaneous fit and errors are computed by Monte Carlo, Izotov \&
Thuan find $0.2565\pm 0.0010\ \mathrm{(stat)} \pm 0.0050
~\mathrm{(syst)}$ from 93 H{\sc ii} regions \cite{izotov10} while Aver
et al.  find $0.2573 \pm 0.0033$ \cite{aver11} for nine selected
low-metallicity objects from the Izotov \& Thuan data set.

The quoted \yhe\ is generally an extrapolation to zero metallicity $Z$
of the helium mass fractions $Y$ observed in objects of (small)
nonzero $Z$.  Peimbert et al. applied a $dY/dZ$ slope calibrated
elsewhere; Izotov \& Thuan determined $dY/dZ$ internally to their data
by linear regression; the Aver et al. result quoted above is the mean
of their nine selected objects, with zero slope assumed.  They also
derived from linear regression with no prior on $dY/dZ$ a result of
$0.2609\pm 0.0117$, but it appears that their narrow metallicity
range does not allow a slope to be determined.  The upper end of this
error interval is close to the initial Pop I composition of the Sun,
$Y_\mathrm{proto} = 0.278\pm 0.008$ \cite{serenelli10}, which must be
greater than \yhe\ since stars are net creators of $^4$He.  We repeat
our analysis for each of these three input \yhe\ constraints.

\subsubsection{$^2$H}

Primordial deuterium is observed in gas of very low metallicity along
our lines of sight to distant quasars.  Quasar light is absorbed by
atomic transitions of neutral hydrogen, and deuterium is observed as
additional absorption with the appropriate isotope shift and doppler
width to be deuterium in the same physical location.  Quasar spectra
that are suitable for inference of D/H are rare, because the quasar
continuum emission must be well-characterized, there must be no other
absorption nearby in redshift/velocity space, and the absorber must
have sufficient hydrogen column density to produce measurable
absorption from its small amount of deuterium.  To date there are
eight extragalactic systems in which D/H is generally agreed to have
been measured.  These are described in
Refs. \cite{pettini08,fumagalli11} (and references therein), where the
average $\log(\mathrm{D/H})=-4.556\pm 0.034$ is found.  As discussed
in Refs. \cite{kirkman03,pettini08,holder10}, it is not clear that all eight
systems have consistent inferred D/H.  We adopt this average as the
observed primordial D/H.

\subsubsection{$^3$He and $^7$Li}

The other two nuclides predicted in measurable quantities by BBN are
$^3$He and $^7$Li.  $^3$He has only been observed in our galaxy (in
H{\sc ii} regions) and only at relatively high metallicity
\cite{bania07}.  Moreover, its post-BBN history is complicated, as it
seems to be produced copiously in some stars but mostly burned up in
others \cite{hogan95}.  Although its observed abundance is generally
consistent with BBN, further inferences from $^3$He seem unwarranted.
The primordial $^7$Li/H has been identified with the nearly-uniform
abundance found in atmospheres of low-metallicity main-sequence
turnoff stars of the Galactic halo
\cite{spite82,asplund06,bonifacio07,fields11}.  As shown in the left
side of Fig. \ref{fig:schramm}, this abundance is near the minimum
producible by standard BBN at any $\Omega_b$ and a factor of 3.5 below
that predicted by the WMAP $\Omega_b$.  Whether this is a consequence
of post-BBN processing (most likely in the observed stars themselves)
or of something missing in standard BBN is unclear \cite{fields11}.
Since we are considering only standard BBN apart from variation of
\neff, the important fact here is that even at $\neff=10$, BBN
overproduces $^7$Li by a factor of 2 relative to halo stars.  Thus, it
is an implicit assumption of our model that the observed Li/H is not
primordial, and we do not include it in our constraints.

\section{BBN and CMB Likelihood Functions}

The cosmic microwave background is affected by extra radiation in two
ways \citep{bashinsky04,hou11}. On large scales the CMB is sensitive
to the gravitational effect of the perturbations in the extra
radiation, while on smaller scales the increased expansion rate leads
to an increased amount of diffusion damping.  However, changing the
helium abundance also affects diffusion damping, as \yhe\ affects the
number of electrons per unit baryon mass.  The recent results of
\citet{keisler11} used CMB power spectra from the WMAP seven-year data
release (WMAP7) and the South Pole Telescope (SPT) to find
$\neff=3.85\pm 0.62$ when the relation between \yhe\ and \neff\ from
BBN theory was enforced, ignoring uncertainties in the BBN rates.
Without an imposed BBN prior on \yhe\ the precision of the
\neff\ measurement is degraded to $\neff = 3.4 \pm 1.0$.  Following
\citet{keisler11}, we use bandpowers from WMAP7 and SPT, which
together cover the multipole range $\ell=2-3000$.

To combine BBN constraints with CMB measurements, we adopt a Markov
Chain Monte Carlo (MCMC) approach, using a modified version of the
publicly available package CosmoMC \cite{lewis02}. MCMC methods work
by sampling the parameter space at random positions and either
accepting or rejecting each sample based on the likelihood of that
particular point in the parameter space.  The most-likely values and
confidence intervals for the parameters are identified from the
distribution of samples.

For CMB measurements, we assume an 8-parameter spatially flat
$\Lambda$CDM cosmology, allowing as parameters the cold dark matter
density ($\Omega_c h^2$), the baryon density ($\Omega_b h^2$), the
angular size of the sound horizon ($\theta_s$), the optical depth to
Thomson scattering in the reionized universe ($\tau$), the amplitude
($A_s$, using as a parameter $\ln A_s$) and spectral index of the
potential fluctuations ($n_s$), the helium abundance (\yhe) and the
effective number of neutrino species (\neff). As additional
parameters, we allowed for a contribution to the SPT power from the
Sunyaev-Zeldovich effect, clustered galaxies, and shot noise from
galaxies, using priors from \citet{shirokoff11}, as described in
\citet{keisler11}. We assume broad uniform priors that extend well
beyond the WMAP7-allowed range on all cosmological parameters and
explore $0<\yhe<1$ and $1.046<\neff<8.046$.

The BBN likelihood only depends on a small subset of these parameters:
$\Omega_b h^2$, \neff, and \yhe. For a given assumed pair of $\Omega_b
h^2$ and \neff, a probability distribution based on nuclear
uncertainties can be calculated for \yhe\ and $D/H$ from BBN theory.
This can be multiplied by the observed joint probability distribution
describing \yhe\  and D/H observations, and the resulting distribution
can then be integrated over $D/H$ to yield a final $P(\yhe)$:
\begin{eqnarray}
P(\yhe, D/H,\neff, \obh|{\yhe}_{obs}, D/H_{obs}) & \propto &   \nonumber  \\
  P({\yhe}_{obs}, D/H_{obs}|\yhe, D/H)    & \nonumber \\
   \times P(\yhe, D/H|\neff,\obh) & &
   \label{eqn:prob}
   \end{eqnarray}
As written, the two factors on the right side of the equation are the
observed abundances and the theory prediction for the abundances, each
of which is a 2-D probability distribution.

The D/H ratio does not affect CMB observables, so we marginalize over it to obtain the final likelihood:
   \begin{equation}
   P(\yhe, \neff, \obh) \propto \int P(\yhe, D/H,\neff, \obh) d(D/H)
   \end{equation}
In general, this procedure could be time-consuming in a MCMC, where
likelihood evaluations may be done hundreds of thousands of times. We
approximate the observed and theory abundances as 2D Gaussians, with
the observed abundances uncorrelated and the (small) theory
covariances extracted from the Monte Carlo BBN calculations described
above. In this case, Eq. (\ref{eqn:prob}) is simply the product of two
Gaussians, the result of which is another Gaussian distribution, and
the marginalization is trivial. Therefore, the likelihood evaluation
and marginalization over the true $D/H$ are reduced to an algebraic
equation involving the means and covariances of the observed and
theory abundances. 

In more detail, following the notation of \citet{ahrendt05}, if the
probability distributions of the theoretical and observational
abundances are given by multivariate normalized normal distributions
$N(\vec{\mu}_{th},\Sigma_{th})$ and $N(\vec{\mu}_{obs},\Sigma_{obs})$,
with means denoted by $\vec{\mu}$ and covariance matrices $\Sigma$,
the product of these two distributions is
\begin{equation}
N(\vec{\mu}_{th},\Sigma_{th}) N(\vec{\mu}_{obs},\Sigma_{obs}) =
z_c N(\vec{\mu}_{comb}, \Sigma_{comb})
\end{equation}
where the combined covariance matrix  is
\begin{equation}
\Sigma_{comb}=(\Sigma_{th}^{-1} + \Sigma_{obs}^{-1})^{-1}
\end{equation}
and the combined mean is 
\begin{equation}
\vec{\mu}_{comb}=\Sigma_{comb}(\Sigma_{th}^{-1}\vec{\mu}_{th} + \Sigma_{obs}^{-1}\vec{\mu}_{obs}).
\end{equation}
The prefactor is given by
\begin{eqnarray}
z_c & = & {1 \over |2\pi(\Sigma_{th}+\Sigma_{obs})|^{-1/2} } \times \\ \nonumber
 & & \exp \Bigl[ -{1 \over 2} (\vec{\mu}_{th}-\vec{\mu}_{obs})^T (\Sigma_{th}+\Sigma_{obs})^{-1}(\vec{\mu}_{th}-\vec{\mu}_{obs}) \Bigr] .
\end{eqnarray}
Marginalizing over D/H is then done by an analytic integral which
leads to the 2-D normal distribution in Equation (3) being simply
replaced by a 1-D normal distribution involving only the mean and
covariance of \yhe.

In summary, to include full BBN information within an MCMC, one must
simply supply observed He and D abundances and a covariance matrix for
the observations (presumably always diagonal), and a grid of BBN
predictions as functions of \obh\ and \neff\ for He and D abundances
with the theoretical uncertainty expressed as a covariance
matrix. While not pursued here, the generalization to include other
abundances (such as $^3$He or $^7$Li) is straightforward in this
framework.

\section{BBN and CMB Constraints on \neff}

Before combining BBN and CMB constraints, it is important to first
check for consistency. CMB-only constraints on $\neff$ marginalized
over cosmological parameters and \yhe\ are weak but still useful, as
can be seen in Fig. \ref{fig:schramm} and Table \ref{table:results}
(and more fully in Fig. 12 of Ref. \cite{keisler11}). Furthermore, we
can use measurements of helium abundances to improve the precision
of the CMB measurement.  This is true even with weak constraints on
\yhe, which can be provided by the protosolar abundance as an upper
limit.

A lower limit on \yhe\ has less impact but also helps.  It has long
been recognized that observed helium abundances everywhere are too
large to be attributed to stars \cite{hoyle64}.  Both observations and
models of chemical evolution find $\Delta Y/\Delta Z$ in the range 1
to 5 \cite{timmes95} at low metallicities and smaller slopes at higher
metallicities where type Ia supernovae contribute to $Z$.  A solar
metallicity of 0.014 \cite{asplund09}, would then imply that stars
have contributed at most $\Delta Y \sim 0.07$.  Since
$Y_\mathrm{proto} = 0.278\pm 0.008$ is the protosolar abundance, this
implies $\yhe \gtrsim 0.22$, consistent with all observations over a
long period of time (cf. Table 1 of Ref. \cite{pagel00} and the
accompanying discussion).

Applying the weak prior that $0.22 < \yhe < Y_\mathrm{proto}$, the
precision of the CMB measurement of \neff\ is improved to $3.87 \pm
0.81$. This is only slightly weaker than constraints on \neff\ using
CMB data combined with the substantially more precise measurements of
\yhe\ from nearby galaxies, which are shown as the first column in
Table \ref{table:results}.

Separately, the more precise determinations of \yhe\ can be combined
with D/H measurements and BBN theory to get a BBN-only measurement of
$\neff$ (and a simultaneously-fit \obh\ that agrees with the CMB),
shown in the last column in Table \ref{table:results}.  The
constraints are formally quite strong, at $\Delta\neff \sim 0.25$, but
results using different \yhe\ determinations inherit the sometimes $>
2\sigma$ disagreement of the underlying \yhe.

A direct comparison of the first and last data columns provides a
non-trivial test of cosmology. For the two higher
\yhe\ determinations, the separate \neff\ values from BBN and the CMB
are in excellent agreement, suggesting that the effective number of
neutrino species was the same in the first few minutes of the early
universe as it was at $\sim$ 300\,000 years. The lower
\yhe\ measurement of Peimbert is in mild tension with this
interpretation, with only a very slight preference for an extra
neutrino species at late times but better agreement with the
expectation that $\neff = 3.046$.

Given the apparent consistency, it is reasonable to combine CMB and
BBN constraints. Simply requiring that the relation between \yhe\ and
\neff\ be consistent with standard BBN theory leads to an improvement
in the CMB measurement, and further combining this constraint with
observed light element abundances leads to the very tight constraints
shown in the second and third data columns. Again, the systematically
different \yhe\ determinations lead to challenges in interpretation:
the combined analysis yields either a clear detection of an excess of
effective neutrinos over the standard value of 3.046 (at
$2.2-3.4\sigma$) or excellent agreement (within $1\sigma$).

\begin{table*}[h]
\begin{tabular}{|c|c|c|c|c|}
\hline
Data            & CMB+\yhe    & CMB+BBN+\yhe  & CMB+BBN+\yhe+D/H & BBN+\yhe+D/H \\ 
\hline
Aver et al      &  $3.82\pm 0.74$ & $3.84\pm 0.24$ & $3.87\pm 0.24$ & $3.85\pm 0.26$ \\   
Izotov \& Thuan &  $3.83\pm 0.75$ & $3.77\pm 0.36$ & $3.83\pm 0.35$ & $3.82\pm 0.45$ \\
Peimbert et al  &  $4.02\pm 0.79$ & $3.18\pm 0.21$ & $3.22\pm 0.19$ & $3.13\pm 0.21$ \\
\hline \hline
CMB only     & \multicolumn{4}{|c|}{$3.4\pm 1.0$} \\
CMB+$0.22<\yhe<Y_\mathrm{proto}$& \multicolumn{4}{|c|}{$3.87\pm 0.81$} \\
CMB+BBN      & \multicolumn{4}{|c|}{$3.89\pm 0.60$} \\
CMB+BBN+D/H  & \multicolumn{4}{|c|}{$3.90\pm 0.44$} \\
\hline \hline
\end{tabular}
\caption{Constraints on \neff\ from various combinations of data sets.
The first three rows of results use three different precise determinations
of \yhe, while the corresponding columns show different combinations
of these \yhe\ measurements with CMB data, BBN theory (including
errors on reaction rates), and D/H abundance measurements. The lower set
of results do not use these precise \yhe\ measurements.}
\label{table:results}
\end{table*}

\begin{figure}
\includegraphics[width=3.4in]{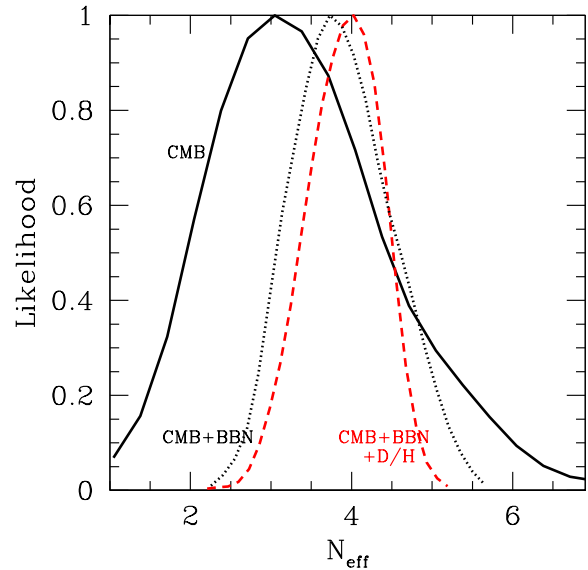}
\caption{(Color online) Constraints on \neff\ from just CMB and $D/H$
  measurements, shown as marginalized likelihoods.  The black solid
  curve is CMB alone, the black dotted curve shows CMB with BBN
  constraints on the relation between \yhe\ and \neff, and the red
  dashed curve also adds the observed abundance of $D/H$.}
\label{fig:dh}
\end{figure}

Given the disagreement between \yhe\ determinations, it is interesting
that a constraint on \neff\ can be obtained using CMB measurements,
BBN theory, and only the D/H abundance, as can be seen in Figure
\ref{fig:dh}.  Using no \yhe\ measurements at all, there is a hint of
excess radiation (at just below $2\sigma$).  This is interesting, as
the current measurement of D/H is at the 10\% level. It is challenging
to find systems that are well-suited for this measurement, but it is
not inconceivable that the D/H abundance measurement could be
sustantially improved.  D/H has long been viewed mainly as a probe of
\obh, but it may have a future as a probe of the expansion rate
encoded by \neff\ with independent systematics, separate from \yhe.

\section{Conclusion}

In this paper, we have presented a comparison of determinations of the
effective number of neutrino species, $\neff$, from recent CMB
measurements (WMAP7 and SPT) and BBN calculations, applying observed
abundances of helium and deuterium in several ways.  Our results are
similar to those of other analyses in the recent literature
\cite{archidiacono11,hamann11-1,gonzalez11,hamann11-2}, though
comparison is difficult because of the use of different cosmological
data and model spaces and the common practice of setting $Y_p=0.24$,
which seems on the basis of Fig. \ref{fig:bbn-variants} to bias
\neff\ a bit high.  Like these and previous analyses, ours generally
favor $\neff > 3$.  The most directly comparable result in the
literature combines BBN and CMB data with constraints on large-scale
structure and the Hubble constant to find 95\% confidence limits of
$\neff = 3.90^{+0.39}_{-0.56}$ \cite{hamann11-1}.  This is similar to
our results.

The BBN calculation used here includes a new estimate of a key cross
section, $d(p,\gamma)^3\mathrm{He}$, based on detailed new theoretical
calculations, which shifts deuterium yields by $\sim 6\%$ and lithium
yields by $\sim 10\%$. This new cross section shifts \obh\ or
\yhe\ inferred from D/H and slightly aggravates the large discrepancy
between standard BBN and observed $^7\mathrm{Li}$ abundances.

The BBN calculations and observed abundances have been embedded in a
module that can be used in a widely used MCMC code (CosmoMC)
and combined with CMB measurements or other cosmological data. 
The BBN likelihood calculation
includes both theoretical and observational uncertainties, including
the weak correlation between computed D/H and \yhe.

We found that there is good agreement between BBN-based and CMB-based
determinations of \neff. These tests probe two very different epochs
of the universe (a few minutes versus a few hundred thousand years),
providing evidence that the same cosmological model describes both
epochs.  However, both probes show hints of new physics in that the
preferred number of effective neutrino species is above the standard
model expectation. This is especially true in the case of the two
higher \yhe\ determinations used (based on overlapping data), with as
much as a $3.4 \sigma$ excess. 

Setting aside the helium constraints and
combining the CMB with BBN theory and the observed D/H, the excess is
$2\sigma$, at $\neff=3.9\pm 0.44$. 
Improvements in the measurement of the D/H abundance would
be extremely valuable as both a cosmological probe and an external
check on \yhe\ determinations.

The results here underscore the complementarity of CMB measurements,
BBN theory, and light element abundance measurements.  CMB
measurements allow both an increased precision in BBN predictions
(through a robust measurement of the baryon density) and a strong test
for BBN (through comparisons with \neff\ and \yhe\ determinations).
This is a consequence of the strong linkages in the standard cosmology
between light element yields in the early universe and observable
signatures at recombination, which connect the underlying cosmological
model, the acoustic oscillations and Silk damping observed in the CMB,
and the abundances observed in low-metallicity systems.  Precise
measurements of all of these interlocking observables provide strong
tests of the model and its parameters.

\begin{acknowledgments}
  
  We acknowledge Ryan Keisler for extensive assistance in using the
  SPT bandpowers, Robert Lopez for providing his independent BBN code,
  Laura Marcucci for providing a machine-readable table of the Pisa
  cross sections for $d(p,\gamma)^3\mathrm{He}$, and Laura Marcucci,
  Rocco Schiavilla, Michael Turner, and R. B. Wiringa for useful
  discussions.  KMN was supported by the U.S. Department of Energy,
  Office of Nuclear Physics, under contract No. DE-AC02-06CH11357; GH
  was supported by NSERC, a Canada Research Chair, and the Canadian
  Insitute for Advanced Research. Calculations were carried out using
  the CLUMEQ computing facilities.
\end{acknowledgments}


\end{document}